\documentclass[article]{revtex4}
\usepackage{pdfpages}
\usepackage{epsf}
\usepackage{subfigure}
\usepackage{amsmath}
\usepackage{eqnarray,amsmath}
\usepackage{epstopdf}
\usepackage{mathrsfs}
\usepackage{natbib}
\usepackage{amssymb,latexsym}
\usepackage{dcolumn}
\usepackage{graphicx}

\def\be{\begin{equation}}
\def\ee{\end{equation}}
\def\ba{\begin{eqnarray}}
\def\ea{\end{eqnarray}}

\graphicspath{{images/}}
\begin{document}
	\title{\large \bf  Long gradient mode and large-scale structure observables}
	\author{Alireza Allahyari}
	\affiliation{Department of Physics, Sharif University of Technology, P.O. Box 11155-9161,	Tehran, Iran }
	\email{allahyari@physics.sharif.edu}
	
	\author{Javad T. Firouzjaee}
	\affiliation{ School of Astronomy, Institute for Research in Fundamental Sciences (IPM), P. O. Box 19395-5531, Tehran, Iran }
	\email{j.taghizadeh.f@ipm.ir}

\begin{abstract}
We extend the study of long mode perturbations to other large scale observables such as cosmic rulers, galaxy number counts and halo bias.
The long mode is a pure gradient mode that is still outside observer's horizon. We insist that  gradient mode effects on observables vanish. It is also crucial that the expressions for observables are relativistic. This allows us to show that the effects of a gradient mode on the large scale observables vanishes identically in a relativistic frame work. To study the potential modulation effect of the gradient mode on halo bias, we derive a consistency condition to the first order in gradient expansion. We  find that the matter variance at a fixed physical scale is not modulated by the long gradient mode perturbations when the consistency condition holds. This shows that the contribution of long gradient modes to bias vanishes in this frame work.
\end{abstract}
%
%

\maketitle
\section{Introduction}

Latest developments in cosmology are provided by deep redshift surveys for cosmological structures. These structures develop from evolution of primordial perturbations
generated during inflation. Originally, quantum fluctuations of the inflation field set the seeds of curvature perturbations at primordial epochs. The nature of these fluctuations are well studied in the Cosmic Microwave Background (CMB). The fingerprints of these fluctuations on large scale structures (LSS) are also important since the current and future galaxy surveys will take high precision measurements of galaxy clustering. This provides enough statistical precision to solve the issues in the standard model of cosmology. The importance of LSS is highlighted by the fact that LSS surveys are three dimensional, whereas the CMB is two dimensional.  We study the fingerprints of long modes that come from inflation using the relativistic approach to large scale observables.\\

To interpret galaxy and LSS survey observations correctly, one should note that as the light propagates through the cosmos, its path is modified by inhomogeneities. Therefore, one needs to use general relativistic calculations to relate the observables to the inhomogeneities. Since the exact relativistic calculation for cosmological structure observables is complicated \cite{exactsolution}, usually the first or the second order perturbations of the Einstein equations are used to calculate these observables. 
Some of these effects on the observed fluctuations of galaxies are the dark matter density fluctuations, the redshift-space distortions (the peculiar velocity effect in the redshift space of light) and the magnification bias which are studied in \cite{ Jeong:2014ufa, Yoo:09-10, Bonvin:11-14, Challinor:2011bk}.
The standard method to derive the full relativistic expression for galaxy clustering is done by tracing back the photon path given the observed redshift and the angular position of the source galaxies. A fully general  relativistic expression for the observed galaxy density contrast includes
volume distortions due to the light deflections, evolving number densities, galaxy bias, as well as the magnification bias generalized to the evolving luminosity function. These  relativistic effects are usually named projection effects.\\

Generally, there are different wavelengths for the gravitational potential that come from the early universe fluctuations. The long wavelength potential can produce a superhorizon perturbation that might have observable effects on the CMB  or on  other large scale structure observables. The power asymmetry in the CMB  as observed by Planck satellite \cite{planck13} (for earlier reports of hemispherical asymmetry in WMAP data see \cite{wmap}) can be the generic property of some early universe models. This issue has been recently revisited in  works such as \cite{Akrami:2014eta} and \cite{Ade:2015hxq} which use a local-variance estimator. The observations show that the power spectrum in the northern hemisphere is different from  the power spectrum in the southern hemisphere. Erickcek et al. \cite{Erickcek:2008sm} have proposed that a superhorizon perturbation would introduce a preferred direction that generates the power asymmetry. Along this way, the predictions of inflationary models with  a long mode modulation of large scale structures are presented in \cite{Namjoo:2014nra}. As a matter of fact, it was shown that a long constant \cite{Weinberg} and a long gradient mode \cite{Hinterbichler:2012nm} can be gauge artifacts of the perturbation theory which does not leave any effect  on the CMB \cite{Creminelli:2011sq, Mirbabayi:2014hda}. Note that the vanishing contribution of a gradient to the CMB is non trivial.
Now, our aim is to study the non trivial effects of long modes which could be possibly observed in the future galaxy surveys. Our goal in this paper is to extend the investigation of the long gradient modes effects to other large scale observables like the galaxy number counts considering the subtleties that arise from the  bias.\\

Given the high precision of LSS surveys and  some hints of the CMB asymmetry from long mode perturbations, we study the imprints of these long modes on  other large scale observables as well. The cancellation of the pure gradient modes in the galaxy number counts is nontrivial as various new effects appear compared to the CMB case. Specifically, we show that the contribution of a pure gradient mode to large scale observables vanishes. Using this frame work, we show that the coupling of the long gradient modes to short modes does not modulate the smoothed matter variance  at a fixed physical scale if some form of consistency condition is satisfied. Therefore, we confirm the consistency of the definition (\ref{halo}). \\

The outline of the paper is as follows. We begin with a review of the long mode effects in the perturbation theory in Section II. We also discuss the importance of  extending the previous studies to other large scale observables. The study of large scale observables is presented in two sections. We calculate the three large-scale observables in the presence of long gradient mode in section III, after reviewing the relativistic derivation of large scale observables. The main section is section IV. It is devoted to the study of the galaxy clustering quantities such as the galaxy number counts and the halo bias in presence of a long gradient mode. Finally, our conclusion and discussion are presented in section V. The Latin indices indicate the space components and Greek indices indicate the space time components and we have set $c=1$.
\section{long gradient mode}

In cosmological perturbation theory, gauge invariant quantities are defined with the assumption that perturbations fall off at infinity. This allows to decompose perturbations as scalars, vectors, and tensors (SVT). Examples of perturbations that do not vanish at infinity are zero momentum modes and pure gradient modes.\\

\textbf{Zero momentum modes:}
In the perturbation theory zero momentum modes are important, because they modulate the power spectrum in the case of single field slow-roll inflation as derived in \cite{Maldacena:2002vr}. This modulation is known as the consistency condition. These zero momentum modes will induce a local type non-Gaussianity. It is argued that the consistency relations are true in models of inflation in which the only dynamical field is the inflaton field \cite{Creminelli:2004yq}. These consistency relations are also derived in the Newtonian limit in \cite{Kehagias:2013yd}. Although this type of non-Gassainity is small and proportional to $n_{s}-1$, it is crucial to know if our large scale observations are really contaminated by such effects. 
 
Along this line, it is shown that even after fixing the gauge,  zero momentum transformations are still allowed \cite{Weinberg}. This remaining gauge freedom is shown to be the source of IR divergences \cite{Urakawa:2010it}. That is because one needs boundary conditions to uniquely solve for the lapse function after imposing the conventional comoving gauge condition. It is shown that zero momentum modes in the metric can be removed by a coordinate transformation, so they do not have any observable effect as they induce a relative shift in the expansion history. The vanishing effect of a constant mode in the power spectrum is emphasized in \cite{Dai:2013kra}.\\

\textbf{Gradient modes:}
Let us suppose that we have a long wavelength perturbation expanded as $\frac{k}{a H}\varphi_{k}$ in the Fourier space, where $a$, $H$ and $k$ are the scale factor, the Hubble parameter and the Fourier mode for the long mode respectively. 
This perturbation appears in the expansion of primordial perturbations of the form $\sin(k.x)$. The average of such perturbations  should vanish in the whole universe in order to keep homogeneity and isotropy. We suppose that the effect of such a perturbation does not vanish in our Hubble patch.
It is also shown that pure gradient modes are removed by a coordinate transformation in the metric \cite{Hinterbichler:2012nm, Mirbabayi:2014hda}. In the Newtonian gauge this transformation is given by
\begin{align}
&\tau \longrightarrow \tau+\epsilon \\ \nonumber
& x^{i} \longrightarrow x^{i}(1+k.x)-\frac{1}{2}k^{i}x^2-f(\tau)k^i,
\end{align}
where $\epsilon$ and $f$ are defined as
\begin{align}
&\epsilon=-f^\prime k.x \\ \nonumber
&f^\prime=\frac{1}{a^2}\int a^2 d\tau.
\end{align}
That is because the effect of a pure gradient mode is an acceleration which is not observable locally.\\

In this manner, we extend the gradient mode studies to other large scale observables including cosmic rulers, galaxy number counts and the halo bias.
The vanishing effect of gradient modes on observables is not trivial given that various relativistic terms appear (equation (\ref{galaxy-p})).
A consistent general relativistic formulation of galaxy clustering should  take into account all effects such as displacements, velocities and other new relativistic effects naturally. The especial case of the halo bias needs to be treated in a different way, because we need to consider coupling to short modes. We will show that in the general relativistic formulation,  the effects of a gradient mode vanish if some forms of consistency condition hold.\\

To study gradient mode effects, one needs to define the observables. This is done in the relativistic frame work for observables. The observables are gauge invariant as a prior and made of information on the past light cone. They should to be devoid of gauge artifacts as well. A long gradient mode is gauge artifact.
The observables are described by the standard clocks, the standard rulers and the galaxy number counts \cite{Jeong:2014ufa}. One example of such a clock is the CMB on large scales. Next section is devoted to study of effects of long gradient modes on observables.\\

The other cosmological observable is the galaxy number count. We also study the gradient mode effects on galaxy number counts. Galaxy number counts importance is highlighted by the fact that future surveys will probe the cosmos on large volumes with unprecedented precision. Because of their significant statistical precision, they will be able to discriminate between inflationary models by the level of non-Gaussianity. It is crucial to consistently formulate all the effects in order to  correctly interpret the data.
Other relativistic formulations have recently emerged \cite{Jeong:2014ufa, Yoo:2014kpa, Bonvin:11-14}. The authors in \cite{Challinor:2011bk} use the evolution equation for  Jacobi map to derive the galaxy number counts.
We found that our result of  the galaxy number counts agrees with the results given in \cite{Kehagias:2015tda}. In next sections, we study the imprints of gradient modes using relativistic definition for large scale observables.

\begin{figure}[htbp!]
	\centering
	\includegraphics[width=0.7 \columnwidth]{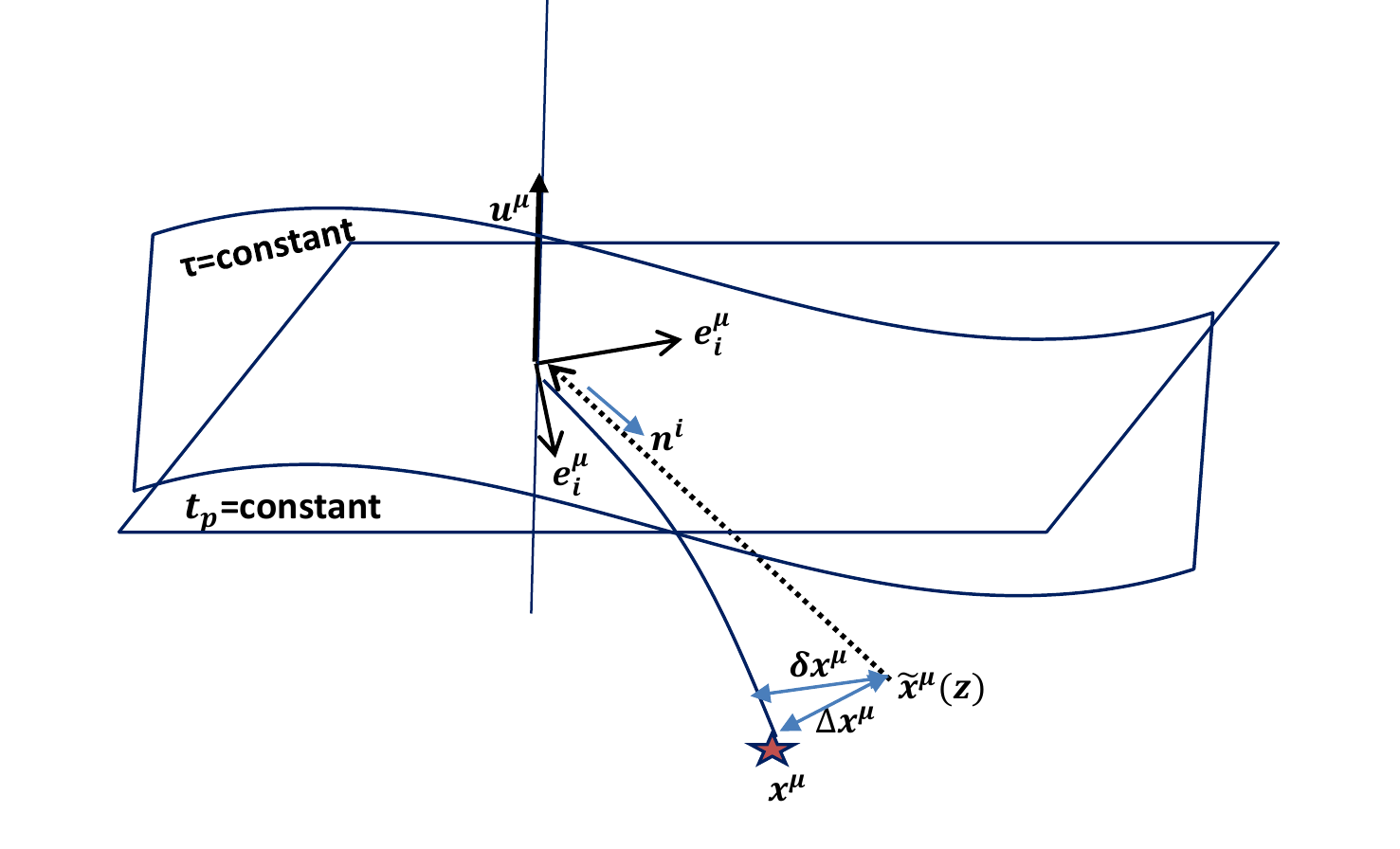}
	\caption{The apparent position of a galaxy is shown relative to its real position for an observer with four velocity $u^\mu$.}
	\label{fig:1}	
\end{figure}

\section{large-scale observables I: cosmic rulers}
One of the interesting observable quantities in the LSS is the galaxy number density fluctuation.
This quantity is measured in the redshift space.  It is well known that the observed power spectrum for galaxy number density is modified by  the peculiar velocities as described by Kaiser formula \cite{Kaiser}. However, for large sky surveys this formula is hindered by the fact that this formula does not include all the relativistic terms consistently. Basically, the problem of deriving the relativistic correction terms starts by tracing photons paths in an inhomogeneous universe. Since photons are traveling through inhomogeneities, their path will deviate from the straight path in the FRW universe and these inhomogeneities change the observed size of rulers, like the baryon acoustic oscillations (BAO) or the power spectrum peak and survey volume.
In addition, on large scales, gauge effects become important and should be taken into account. Hence, a gauge invariant formulation is important. For example, galaxies are thought to trace the underlying dark matter overdensity and are biased by a scale independent factor $b_{g}$ with respect to the dark matter overdensity. This definition is obscured because changing the gauge will introduce scale dependencies. In addition, observables should have physical interpretation. We will first review how the relativistic effects are derived. The details of derivations can be found in \cite{Jeong:2014ufa,Yoo:2014vta}. We only give the main points of derivation here to introduce the notation. We will then show that the gradient mode contribution to all cosmic ruler perturbations vanishes identically.\\

\subsection{Photons in an inhomogeneous universe}
An observer screen space is a two dimensional space orthogonal to his velocity $u^{\alpha}$ and the photon four-vector $p^{\mu}$. The observer sees  the cosmos images by projecting on to his own instantaneous screen space
spanned by orthonormal basis ${\tilde{e}}_{1}$ and ${\tilde{e}}_{2}$. These are also orthogonal to the observer's velocity \cite{ellis book}. \\

When a photon is observed at the redshift $z$ (observed redshift), its apparent position is attributed as $\tilde{x}^{\mu}(z)$. In this frame work, the apparent position is always computed assuming a homogeneous universe.
\footnote{$\tilde{} $ refers to apparent quantities.}. Note that $\tilde{x}^{\mu}(z)$ also depends on
direction at the observer.
The photon trajectory at the observed redshift $z$ is deviated from a similar trajectory in a homogeneous universe. The actual position of the source is obtained by integrating the geodesic equations for the photons in an inhomogeneous universe as shown in  Fig.(\ref{fig:1}). These deviations are gauge dependent and unobservable.
Because null geodesics are conformally invariant, it is simpler to use conformally transformed metric and affine parameter for the photons. In this way the photon geodesics have simple expressions. But it is important to use the correct metric in the expressions (\ref{proj1}, \ref{redshift}).   The photon geodesics  is solved for the source position in \cite{Jeong:2014ufa, Yoo:09-10}. We present the final result as 
\begin{align}
x^{\mu}=\tilde{x}^{\mu}(z)+\Delta x^{\mu}=\tilde{x}^{\mu}(z)+\delta x^{\mu}+\frac{d{\tilde{x}}^{\mu}}{d\chi}\delta\chi,
\label{res}
\end{align}
where $ \chi$ is the conformal affine parameter for null geodesics in the background and $\tilde{x}^{\mu}(z)$ is apparent position evaluated at the observed redshift $$\tilde{x}^{\mu}(z)=(\tau_{o}-\chi (z),n^{i}\chi (z)).$$ Please note that ($\:\tilde{}\:$) is used for apparent values of any quantity and apparent values are evaluated using the background equations. The first term is given by the geodesic equation (\ref{geo-eq1}, \ref{geo-eq2}) and the second term is given by the redshift matching equation (\ref{redshift-eq}). The perturbed conformal photon four-vector is also $k^{\mu}=(-1+\delta \nu, n^{i}+\delta n^{i})$. The metric in a general gauge is written as $$ds^2=a^2(-(1+2A)d\tau^2-2B_{i}d\tau dx^{i}+(\delta_{ij}+h_{ij})dx^i dx^{j}).$$

Typically, the observer fixes the scale factor at  his observation time by $a(t_{p})=1$, while the global scale factor is simply $a(\tau_{o})\neq 1$ where $t_{p}=\int_{0}^{\tau_{0}} \sqrt{-g_{00}(x_{o},\tau)} d\tau$ is the observer's proper time and the subscript $o$ refers to the observer. 
Note that to first order, proper time is not affected by the peculiar velocity of
the observer. We use the conformal Newtonian gauge in which the metric is given by
\begin{align}
ds^2=a^2(-(1+2\varphi)d\tau^2+(1-2\psi)dx^i dx_{i}).
\end{align}
In the matter dominated Einstein-de Sitter (EdS) universe where there is no anisotropic stress,  we have $\varphi=\psi$. In this frame work,
the scale factor difference between the global scale factor $a(\tau_{o})$ and the $a(t_{p})$ is
\begin{align}
\delta a=a(\tau_{o})-a(t_{p})=-H_{o}\int_{0}^{\tau_{o}}\varphi(x_{o},\tau) a(\tau) d\tau.
\label{a-def}
\end{align} 
Because this expressions are correct up to first order, $a$ refers to the global scale factor in our notation.
Before discussing cosmic rulers, it is helpful to  mention the main points of deriving the equation (\ref{res}). 
As the observations are in observer's rest frame, the initial conditions by requiring that in observer's frame the frequency and the direction of the photons are simply 
\begin{align}
\label{proj1}
1=(a^{-2}g_{\mu\nu}e^{\mu}_{0} k^{\nu})_{o},\\ \nonumber
n_{i}=(a^{-2}g_{\mu\nu}e^{\mu}_{i}k^{\nu})_{o},
\end{align}
where $e^{\mu}_{\nu}$ are the orthonormal tetrads given by $e^{\mu}_{0}=u^{\mu}=a^{-1}(1-\varphi,v^i)$ and $e^{\mu}_{i}=a^{-1}(v_{i},\delta_{i}^{j}+\psi\delta_{i}^{j})$ where $i$ is the tetrad index and $j$ refers to space time indexes. In tetrads definition, $v_{i}$ refers to the conformal velocity of the observer. Note that the factor $a^{-2} $ appears as we need the physical photon four-vector $a^{-2}k^{\mu} $ in order to project on the observer's frame. In this way, the perturbation at observer's position is fixed. 
\footnote{For any vector $v^{i}$ we have defined $v_{\lVert}=v^{i}n_{i}$ and $v^{i}_{\perp}=\mathcal{P}^{ij}v_{j}$ in which $\mathcal{P}^{ij}=\delta^{ij}-n^{i}n^{j}$ is the projection operator.}
 Given the initial conditions, Integration of the geodesic equations yield temporal and spatial displacements \cite{Jeong:2014ufa}
\begin{align}
\label{geo-eq1}
\delta x^{0}=&(-\delta a_{o}-\varphi_{o}-v_{||o} )\chi+\int \left(  2\varphi+(\chi-\chi^{\prime})(\dot{\varphi}+\dot{\psi}) \right)d\chi^{\prime}-\int_{0}^{\tau_{o}} \varphi(x_{o},\tau)ad\tau ,\\
\delta x^i=&(\delta a_{o} n^i-\psi_{o}+v_{o}^i)\chi +\int \left(\psi n^i+(\chi-\chi^{\prime})(-\partial_{i}\varphi-\partial_{i}\psi) \right)d\chi^{\prime},
\label{geo-eq2} 
\end{align}
where $\delta a_{o}=a_{o}-1$ and dot denotes derivative with respect to the conformal time $\tau$. Note that equation (\ref{a-def}) is also used. We stress that that $\chi$ is calculated at the observed redshift. This shifts the real value of $\chi$ by $\delta \chi$. The value of $\delta \chi$ is derived by redshift matching. The redshift for the photons is defined by
\begin{align}
1+z\equiv \frac{1}{\tilde{a}}=\frac{(a^{-2} k_{\mu}u^{\mu})_{e}}{( a^{-2} k_{\mu}u^{\mu})_{o}}=\frac{(1+\varphi+v_{\lVert}-\delta\nu)_{e}}{a(x^{0})},
\label{redshift}
\end{align}
where $x^{0}$ is the real emission time.
The scale factor perturbation with respect to the apparent scale factor $\tilde{a}$ is defined as $\Delta \ln a=\frac{a}{\tilde{a}}-1$. This is given by 
\begin{align}
\Delta \ln a=\frac{\partial \ln a}{\partial \tau}(x^{0}-\tilde{x}^{0}).
\end{align}
The equation (\ref{redshift}) yields \cite{Jeong:2014ufa}
\begin{align}
\delta\chi=\delta x^{0}-\frac{1+z}{H}\Delta \ln a.
\label{redshift-eq}
\end{align}.
Now that we have reviewed the notation and main points, we move on to the observables in the next section.

\subsection{Cosmic rulers}

The cosmic rulers are the cosmological observables that their spatial scale $r_{0}$ is known like the CMB on large scales. Because we know the scale when photons decoupled, in principles, this scale can be measured in a statistical manner. Other cosmic rulers include the known scales in the matter power  spectrum or  the baryon acoustic oscillations (BAO) \cite{Fabian}. Moreover, the angular diameter distance perturbations for cosmic rulers are related to the luminosity distance perturbations by $\frac{\Delta D_{L}}{D_{L}}=-\frac{1}{2}\mathcal{M} $ \cite{Bonvin:2005ps} (equation (\ref{M})). These are used for studying the nature of dark energy \cite{Barausse:2005nf}.
The fully relativistic approach to cosmic rulers can be found at \cite{Jeong:2014ufa}. Here we review the main points.
 Basically, the size of cosmic rulers which is measured is their apparent size. The cosmic rulers size change from $r_{0}$ to the observed apparent size $\tilde{r}$.
Because length is a frame dependent quantity, one needs to define the observer who actually measures the length of the rulers. One natural way is to define their length in the instantaneous rest frame of the comoving observers. The velocity of comoving observers is given by $v^i=\frac{T^{i}_{0}}{\rho+p}$ and their rest frame metric is simply $(g_{\mu\nu}+u_{\mu}u_{\nu})$. Provided that the size of the ruler is small, its apparent length yields
\begin{align}
\tilde{r}^{2}=\tilde{a}^{2}(z)\left( -(\delta \tilde{x}^{0})^2+\delta_{ij}\delta\tilde{x}^{i}\delta\tilde{x}^{j}\right), 
\end{align}
whereas in the observer rest frame, its physical length is
\begin{align}
r_{0}^{2}=(g_{\mu\nu}+u_{\mu}u_{\nu})\left(\delta\tilde{x}^{\mu}+\Delta x^{\mu}-\Delta x^{\prime \mu} \right)\left( \delta\tilde{x}^{\nu}+\Delta x^{\nu}-\Delta x^{\prime \nu}\right),  
\end{align}
where $\delta \tilde{x}^{u}=\tilde{x}^{\mu}-\tilde{x}^{\prime \mu} $ is the apparent distance between to points of the ruler  and $\Delta x^{\mu}$ is given by the equation (\ref{res}).
Finally, the relative ruler perturbation is defined as
\begin{align}
\frac{\tilde{r}-r_{0}}{\tilde{r}}=\mathcal{C}\frac{(\delta \tilde{x}_{\lVert})^2}{\tilde{r}^{2}_{c}}+\mathcal{B}_{i}\frac{\delta \tilde{x}_{\lVert}\delta \tilde{x}^{i}_{\perp}}{\tilde{r}_{c}^{2}}+\mathcal{A}_{ij}\frac{\delta \tilde{x}^{i}_{\perp}\delta \tilde{x}^{j}_{\perp}}{\tilde{r}^{2}_{c}},
\label{rulers}
\end{align}
where $\tilde{r}_{c}\equiv \frac{\tilde{r}}{a}$ \cite{Jeong:2014ufa}. To decompose the apparent separation $  \delta \tilde{x}_{i}$ to parallel and perpendicular to line of sight components, the projection operator $\mathcal{P}_{ij}=\delta_{ij}-n_{i}n_{j}$  
is introduced. That is the perpendicular components of separation $\delta\tilde{x}^{\mu}$ are defined as $\delta \tilde{x}^{i}_{\perp}=\mathcal{P}_{ij}\delta \tilde{x}^{j}$. The parallel component is simply given by $\tilde{x}_{\lVert}=n_{i}\delta \tilde{x}^{i}$ \cite{Fabian} and note that the definitions of $\mathcal{A}$,
$\mathcal{B}$ and $\mathcal{C}$ will be given below.
We move on to study the effect of a pure gradient mode on these observables and show that a pure gradient gives vanishing contribution to these observables. This confirms that consistency of this relativistic approach.

\subsubsection{2-scalar  $\mathcal{C}$ }
The scalar $\mathcal{C}$  includes the line of sight perturbations of the cosmic rulers and induces the perturbations in the redshift space. It also includes the redshift space distortion term which is the dominant term on small scales. The first term $\mathcal{C}$  is given by
\begin{align}
\mathcal{C}=-\Delta \ln a+\psi-v_{\lVert}-\partial_{\chi}\Delta x\lVert, 
\end{align}
 \cite{Fabian}.\\

To address the problem of consistency of derivation, we calculate the effect of a pure gradient mode on  $\mathcal{C}$. In the EdS universe a pure gradient mode is $\psi=\phi=k.x$. In the Newtonian gauge,  the scalar $\mathcal{C}$  yields
\begin{align}
\mathcal{C}&=-\Delta\ln a \left( 1-H(z)\frac{\partial}{\partial_{z}}(\frac{1+z}{H})\right)-\varphi-v_{\lVert}+\frac{1+z}{H(z)}(-\partial_{\lVert}\varphi-\dot{v}_{\lVert}) \\ \nonumber
&=\left( +\frac{2}{3}\chi(\frac{3}{2})-\chi+\frac{\sqrt{a}}{H_{o}}(-\frac{1}{3}+\frac{2}{3H_{o}\tau_{o}})\right) k_{\lVert}=0,
\label{C}
\end{align}
where $v_{\lVert}=-\frac{2}{3}\frac{\sqrt{a}k_{\lVert}}{H_{o}}$ and $k_{\lVert}=k_{i}n^{i}$s. We use the fact that $\partial_{\lVert} v^{i}=n^{j}\partial_{j} v^{i}=0$ and
$$\Delta \ln a=v_{\lVert}-v_{o}-\varphi=-\frac{2}{3}\chi k_{\lVert},$$ 
in the first line. We also use $\chi=\frac{2}{H_{o}}(1-\sqrt{a})$. Please note that in the last line we need $H_{o}\tau_{o}$ to zero order. That is we have $a(t_{p})\simeq a(\tau_{o})= 1$ for the normalization of scale factor. This means that $H_{o}\tau_{o}=2$.
As a result, there is consistency in derivation of the line of sight perturbations.

\subsubsection{2-vector $\mathcal{B}$}
The second term  $\mathcal{B}^{i}$ is basically vector. This term is determined by
\begin{align}
\mathcal{B}_{i}=-v_{\perp i}+\frac{1+z}{H(z)}\partial_{\perp i}\Delta\ln a,
\end{align}
where $\partial_{\perp i}={P}_{i}^{j}\partial_{j}=(\delta_{i}^{j}-n_{i}n^{j})\partial_{j}$ \cite{Fabian}.
This term produces perturbations both in the line of sight and perpendicular to the line of sight.
 This terms can be written in the spin basis as
\begin{align}
_{\pm1}\mathcal{B}=m^{i}_{\mp}\mathcal{B}_{i}=-v_{\pm}+\frac{1+z}{H}\partial_{\pm}\Delta \ln a,
\end{align}
where $\textbf{m}_{\pm}=\frac{\textbf{e}_{1}\mp i\textbf{e}_{2}}{\sqrt{2}}$ and $\textbf{e}_{1}$ and $\textbf{e}_{2}$ are the basis on the sphere.\\

To study the effect of the gradient mode, a simple calculation yields $\mathcal{B}_{i}=(\frac{2}{3H_{0}}-\frac{2}{3H_{0}})\sqrt{a}k_{\perp i}=0$.
Again the effect of the pure gradient to $\mathcal{B}_{i}$ vanishes identically. Similar to the former case, the $\mathcal{B}$ term is not modified by the pure gradient mode.
\subsubsection{Magnification and shear}
The last term $\mathcal{A}_{ij}$ in the equation (\ref{rulers}) is the most crucial term, since this term includes we observe in shear surveys. This term is given in a general gauge as \cite{Fabian}
\begin{align}
\mathcal{A}_{ij}=-\Delta \ln a\mathcal{P}^{i}_{j}-\frac{1}{2}\mathcal{P}_{i}^{k}\mathcal{P}_{j}^{l}h_{kl}-\partial_{\perp ( i}\Delta x_{\perp j)}-\frac{1}{\tilde{\chi}}\Delta x_{\lVert}\mathcal{P}_{ij}.
\end{align}
The magnification is the trace part of $\mathcal{A}_{ij}$. In conformal Newtonian gauge, Magnification is 
\begin{align}
\mathcal{M}=\mathcal{P}^{ij}\mathcal{A}_{ij}=-2\Delta \ln a -\frac{1}{2}(-4\varphi)+2\hat{k}-\frac{2}{\chi}\Delta x_{\lVert},
\label{M}
\end{align}
where $\hat{k}=-\frac{1}{2}\partial_{\perp i}\Delta x^{i}_{\perp}$ and $\mathcal{P}^{ij}=\delta^{ij}-n^{i}n^{j}$. Note that lensing convergence ($\hat{k}$) is not gauge invariant. Hence, it is not observable. It has been shown that the magnification is not perturbed by gradient modes \cite{Fabian}. The trace free part of the $\mathcal{A}_{ij}$ term is called shear, $\gamma _{ij}$. The shear term can be written in the spin basis as $_{\pm2}\gamma=m^{i}_{\mp}m^{j}_{\mp}\gamma_{ij}$.
In the Newtonian gauge in the spin basis the shear term  yields
\begin{align}
_{\pm2}\gamma=\int(\tilde{\chi}-\chi)\frac{\chi}{\tilde{\chi}}m^{i}_{\mp}m^{j}_{\mp}\partial_{i}\partial_{j}(2\varphi)d\chi.
\end{align}
Since the shear involves the second derivatives of the metric, it is not modulated by the gradient mode. Consequently, surveys like weak lensing surveys which measure the shear are not contaminated.

\subsubsection{Luminosity distance}
The magnification which produces area perturbations is measured in the lensing surveys and surveys which probe the luminosity like supernova surveys. It is related to angular and luminosity distance by $\frac{\Delta D_{A}}{D_{A}}=\frac{\Delta D_{L}}{D_{L}}=-\frac{1}{2}\mathcal{M}$ \cite{Bonvin:2005ps}.
Therefore, the pure gradient will also induce no asymmetry in standard candle observations since its contribution to the magnification vanishes.

\section{large scale observables II: galaxy number counts and bias}
One of the important observable quantities in the large scale structure surveys is the galaxy number count. Similarly, we stress that the effect a gradient mode should vanishes naturally in a consistent derivation. Along this line, we also study the bias. Since one observes tracers and not the underlying matter field,  a general relativistic definition for bias is essential. One problem is the correct gauge to define bias. It is argued that in the synchronous gauge the conventional definition for bias is correct \cite{Donghui}. This is in contrast to the advocated gauge  in \cite{Yoo:09-10} that chooses constant redshift gauge for the bias.  The synchronous gauge is also proposed for a second order calculation of the galaxy clustering \cite{Yoo:2014vta}.\\

Suppose we have a comoving source with four velocity $ u^\mu$. In the instantaneous source rest frame, a volume element is defined by $dV_{\mu}= \sqrt{-g} \varepsilon_{\mu\nu\alpha\beta}dx^{\nu}dx^{\alpha}dx^{\beta}$. The number of galaxies expressed in the observed coordinates is
\begin{align}
N(x)=\int\sqrt{-g} n_{g}^{p} u^{\mu}\varepsilon_{\mu\nu\alpha\beta}\frac{\partial x^{\nu}}{\partial{{\tilde{x}}^{1}}}\frac{\partial x^{\alpha}}{\partial{{\tilde{x}}^{2}}}\frac{\partial x^{\beta}}{\partial{{\tilde{x}}^{3}}}d^3{\tilde{x}}. 
\end{align}
where $n_{g}^{p}$ is the proper number density of the galaxies. This approach to relativistic galaxy number counts is used by \cite{ Jeong:2014ufa, Yoo:09-10, Bonvin:11-14}. We know that the number of galaxies in real space is equal to the number of galaxies in redshift space. This yields
\begin{align}
N(x)=\int (1+\varphi-3\psi)a^{3}\bar{n}_{g}^{p}(z)(1+\delta_{g}^{p}(z,\tilde{x}))[(1-\varphi)\arrowvert\frac{\partial x^i}{\partial \tilde{x}^j}\arrowvert+v_{\lVert}]d^{3}\tilde x=\int \tilde{a}^{3}\bar{n}_{g}^{p}(z)(1+\delta_{g}(z,\tilde{x}))d^{3}\tilde{x},
\end{align}
where $\delta_{g}^{p}(z,\tilde{x})$ is the number density perturbation and $x^{i}=\tilde{x}^{i}(z)+\Delta x^{i}$. Note that it is assumed that $\bar {n}_{g}(z)=\bar{n}_{g}^{p}(z)$.
Thus, the observed number of the galaxies in the observed coordinates is given by 
\begin{align}
\delta_{g}(z,\tilde{x})=\delta_{g}^{p}(z,\tilde{x})-\psi+v_{\lVert}+\partial_{\lVert}\Delta x_{\lVert}+\frac{2}{\chi}\Delta x_{\lVert}-2\hat{k}=\delta_{g}^{p}(z,\tilde{x})+\delta V
\label{galaxy-p},
\end{align}
where $\hat{k}=-\frac{1}{2}\partial_{\perp i}\Delta x^{i}_{\perp}$ and $\delta V=-\psi+v_{\lVert}+\partial_{\lVert}\Delta x_{\lVert}+\frac{2}{\chi}\Delta x_{\lVert}-2\hat{k}$. This formula is the relativistic generalization of Kaiser formula which includes the new relativistic corrections. These corrections are dominant on  large scales where general relativity and other models deviate from each other. Basically, these corrections should be considered in the future surveys to interpret the observations. In this frame work, we show that  pure gradient mode modifications vanish non trivially. On the other hand, on large scales, observations are constrained by the cosmic variance. Detectability of these new relativistic corrections is discussed in \cite{Yoo-test}. The method proposed to overcome cosmic variance is given by multi tracer method which uses the fact that different biased tracers trace the same underlying density field \cite{Seljak}.

 Consider the case of a matter dominated universe with sources and observers comoving with the cosmic fluid. In the presence of the long mode, we have $v^i=-\frac{2}{3}\frac{\sqrt{a}k^i}{H_{o}}$ and $v_{o}^{i}=-\frac{2}{3}\frac{k^i}{H_{o}}$.
In the Newtonian gauge each term in equation (\ref{galaxy-p}) is given by
\begin{align}
\partial_{\lVert}\Delta x_{\lVert}=(\frac{5}{3}\chi-\frac{\sqrt{a}}{H_{o}}(-1+\frac{2}{3H_{o}\tau_{o}}))k_{\lVert}\\
\frac{2}{\chi}\Delta x _{\lVert}=(2\chi+\frac{4\sqrt{a}}{3H_{o}})
k_{\lVert}=(\frac{4\chi}{3}+\frac{4}{3H_{o}})k_{\lVert},
\end{align}	
The equation (\ref{galaxy-p}) yields
$$
\delta V=0.
$$
Consequently, the long mode does not modify  the number count observations. This confirms the consistency of the relativistic derivation.\\

In addition, as we observe galaxies and not the underlying dark matter density field, we need to know the halo bias to relate $\delta_{g}^{t_{p}}$ in the synchronous gauge to the dark matter perturbation $\delta^{t_{p}}$. The other complexity is that we observe on constant redshift surfaces. Given the difference between constant time surfaces and constant redshift surfaces, number density perturbation yields $ \delta_{g}^{t_{p}}=b\delta^{t_{p}}+\frac{d \ln n_{g}}{d \ln a}\delta \ln a=b\delta^{t_{p}}+\frac{d \ln n_{g}}{d \ln a}\mathcal{T}(\boldmath n)$. The $\mathcal{T}(\boldmath n)$ is the cosmic clock perturbation \cite{Jeong:2013psa}. It is important to note that the cosmic clock perturbation by a gradient mode vanishes as 
\begin{align}
\mathcal{T}(\boldmath n)=-\frac{1}{3} \varphi+v_{\lVert}-v_{o}=\frac{1}{3}(\frac{2}{H_{o}}-\frac{2a^{1/2}}{H_{o}}-\chi)k_{\lVert}=0,
\end{align}
where we have $\chi=(\frac{2}{H_{o}}-\frac{2a^{1/2}}{H_{o}})$ and it is assumed that observers are comoving with the cosmic fluid $v_{\lVert}=-\frac{2}{3}\frac{\sqrt{a}k_{\lVert}}{H_{o}}$.  

\subsection{Halo bias}
The aim in this section is to show that a gradient mode does not contribute to the halo bias (equation (\ref{halo})) if some form of consistency condition (equation (\ref{ps})) is satisfied. In other words, gradient modes in single field inflation models do not induce non-gaussian corrections to the bias.  It is known that the bias in presence of local non-guassianities is modulated \cite{wands}. This modulation increases as the scale increases. This is contrary, as equivalence principle requires that effects of very long modes on local scales  vanish 

A correct bias modulation naturally includes equivalence principle. In addition, General relativity can also induce local non-gaussianities because it couples long modes to shorts modes \cite{Bruni:2014xma, Bertacca:2015mca}. It is then very crucial to discriminate between primordial non-gaussianities and gravity induced non-gaussianities. Our goal is to use the gradient mode to argue that bias definition given by equation (\ref{halo}) . What we do is to circumvent second order calculations by using local transformations. To this end, we first derive the consistency condition for a gradient mode.\\ 

In this section we use  the $\zeta$ gauge. Constant time hypersurfaces have the same proper time in this gauge. The spatial part of the metric in this gauge is given by 
\begin{eqnarray}
a(\tau)^{2}(1+2\zeta_{s})\delta_{ij}d\tilde{x}^{i}d\tilde{x}^{j}.
\end{eqnarray}
A gradient mode $k.x$ can be produced by a coordinate transformation of the form
\begin{eqnarray}
\tilde{x}^{i}=x^{i}(1+k.x)-\frac{1}{2}k^{i}x^2.
\label{cor}
\end{eqnarray}
Note that this transformation does not change the gauge. The metric takes the new form
\begin{eqnarray}
ds^2=(1+2\zeta_{s}+2\zeta_{l})\delta_{ij}dx^{i}dx^{j},
\end{eqnarray}
where $\zeta_l=k.x$. In this metric, finding long-short effects needs a second order treatment. Note that we neglect terms of second order in $k$ because we only study long modes to short modes couplings. In other words, if we have a metric of the form
\begin{eqnarray}
ds^2=(1+2\zeta_{l})\delta_{ij}dx^{i}dx^{j},
\end{eqnarray}
the transformation $\tilde{x}^{i}=x^{i}(1-k.x)+\frac{1}{2}k^{i}x^2$ will change the metric to a flat FRW only locally in patches where $k.x\ll 1$. A different approach used for constant modes is to use a local transformation, named conformal Fermi coordinates, adapted to long modes \cite{Pajer:2013ana}. This locally changes the metric to a FRW metric. The extent of validity of our transformation and conformal Fermi transformation are the same.  \\

In the spherical collapse model, halos are formed when the over density of the collapsing  regions exceeds a threshold value $\delta_{c}$.
The perturbation in the number density of halos, $\delta_{h}$, is biased by the presence of the long modes perturbations because it changes the threshold to
$\delta_{c}-\delta_{l}$. In Press-Schechter method the number density of halos depends on the height parameter $\nu=\frac{\delta_{c}-\delta_{l}}{\sigma_{R}}$
where  $\sigma_{R}$ is the smoothed matter variance \cite{Halo} (some more discussion is in this paper draft \cite{Allahyari:2015mwa}).
If there is no coupling of long modes and short modes, the halo number density is simply modeled by
$\delta_{h}=b\delta_{m}$. However, gravity couples long modes to short modes. In this case, the halo number density perturbation is also a function of the long  mode $\zeta_{l}$. One model which includes the effects of long modes to short modes couplings is
\begin{align}
\delta_{h}(\delta_{m},\sigma_{R})=b\delta_{m}+\frac{d\ln n_{h}}{d\ln \sigma_{R}}\frac{d\ln \sigma_{R}}{d\ln \zeta_{l}}\zeta_{l},
\label{halo}
\end{align}
where  $\sigma_{R}$ is the smoothed matter variance at the scale $R$ \cite{Tobias}. Using the equation (\ref{halo}), we confirm that the bias is not modulated by gradient modes. First, we need to know how the gravitational coupling of long modes to short modes  modulate the power spectrum to understand how the bias changes.\\
 
Here, we derive the power spectrum with short modes and long modes coupling corrections. We first write the power spectrum in local coordinates $\tilde{x}^{i}$ and then transform this back to $x^{i}$ coordinates.
The correlation function for the matter perturbation in the $\tilde{x}$ coordinate where the pure gradient mode is removed is written as
\begin{eqnarray}
\xi(\tilde{x})|_{\zeta_{l}}=\int d^3q' e^{iq'.\tilde{x}}P(q'),
\end{eqnarray}
where $\xi(\tilde{x})|_{\zeta_{l}}$ means the correlation function in the presence of the gradient mode. 
After substituting the equation (\ref{cor}) and integration by  parts, we find that the power spectrum in Fourier space is modulated as
\begin{eqnarray}
P\longrightarrow P(1-\frac{d\ln (q^3P)}{d\ln q}(k.x)+\frac{1}{2}(k.x)+\frac{(k.q)(q.x)}{2q^2}\frac{d\ln P}{d\ln q}).
\label{ps}
\end{eqnarray}
This is the consistency condition for gradient modes which is not derived in \cite{Pajer:2013ana} as we have taken into account terms of order
$k_{l}/k_{s}$ in the consistency condition. Note that this result is essentially second order.\\

The power spectrum modulation (equation \ref{ps}) is because of the left freedom to define new coordinates locally after gauge fixing. Here, we show that these effects should vanish in a different way compared to \cite{Pajer:2013ana}. In this way, we measure the variance in a the same scale in local coordinates. That is our local rulers are also perturbed. This is justified as halo formation is a spherical collapse which is a local process. We will show that $\sigma(R)$ does not change in a fixed physical scale. This approach and \cite{Pajer:2013ana} are similar in the sense that the coordinate transformations change the metric to a flat FRW metric locally. \\

The smoothed matter variance is defined as
\begin{eqnarray}
\sigma^2 (R)=\int d^3q^\prime W^{2}_{R}(q^\prime R) P(q^\prime),
\end{eqnarray}
where $W_{R}(q'R)$ is the window function and $P(q')$ can be substituted from the equation (\ref{ps}). The length scale $R$ is the proper length in $x^{i}$ coordinates. This variance is measured in $x^{i}$ coordinates where the gradient mode is present. Fixing the physical scale, using local metric to measure the radius, is equivalent to a transformation in Fourier space. The transformation from to $q^\prime$ to $q$ is given by
\begin{align}
q^{\prime i}=q^i+q^i(k.x)-\frac{1}{2}(k.x)x^i.
\label{F-eq}
\end{align}
This transformation keeps the extent of the window function the same.
Doing this transformation yields
\begin{eqnarray}
\sigma^2(\tilde{R})=\int d^3q|\frac{\partial q^\prime}{\partial q}|W_{\tilde{R}}^2(q\tilde{R})P(q^i+q^i(k.x)-\frac{1}{2}(k.x)x^i),
\end{eqnarray}
where $|\frac{\partial q^\prime}{\partial q}|$ is the Jacobin of the transformation. The new radius $\tilde{R}$ which is now measured in $\tilde{x}^{i}$ coordinates, is basically where spherical collapse calculations are done. The $\tilde{R}$ is calculated keeping $q.x$ the same after using the equation (\ref{F-eq}). Note that in the integral one can replace $x^{i}$ with $\tilde{x}^{i}$ to first order in $k$.
Finally, the matter variance is 
\begin{eqnarray}
\sigma^2(\tilde{R})=\int  d^3q(1+3(k.x)-\frac{1}{2}(k.x))(1+\frac{d\ln (q^3P)}{d\ln q}(k.x)-\frac{(k.q)(q.x)}{2q^2}\frac{d\ln P}{d\ln q})W_{\tilde{R}}^2(q\tilde{R})P(q).
\end{eqnarray}
This will precisely cancel the change in the power spectrum which we previously calculated in the equation (\ref{ps}). Note that the halo number density depends on the matter variance. Since we showed that $\sigma(\tilde{R})$ is not modified, the equation (\ref{halo}) is nominated by general relativity considerations.\\

As we discussed, the fact that long constant modes can be produced by coordinate transformations is due to consistency condition. However, They will produce a squeezed bi-spectrum by modulation of power spectrum  through gravitational coupling in the coordinates where long mode is present. 
Because we derived a consistency condition for the gradient modes in terms of modulation of the power spectrum, we expect their effect vanish in local coordinates. Here, we use a different transformation than \cite{Pajer:2013ana} and required the same physical scale in local coordinates.
As halo formation is a local phenomena we argue that our transformation is sufficient for capturing the local effects.\\

On large scales, there are other complexities since gauge effects become important. Changing the gauge will produce scale dependencies in the bias. Thus, one needs to find a relativistic definition for the bias. The synchronous gauge is shown to be the right gauge \cite{Donghui}. Since halo formation is a local phenomena, it only depends on the dark matter perturbations in this gauge. This gauge has been implemented for writing the relativistic galaxy clustering. A relativistic bias has also been defined using Fermi normal coordinates and it is shown that it reduces to the usual bias on  small scales \cite{Tobias}.\\

\subsection{Galaxy power spectrum}
Finally, we compute the effect of the pure gradient mode on the observed power spectrum. We can compute the intrinsic power spectrum in a chosen gauge and transform it back to the redshift space using the equations in \cite{Pajer:2013ana}. This equation is given by
\begin{align}
\tilde{\xi}(\tilde{r},z)=\left(1-a_{ij}\tilde{x}^{i}\partial^{j}+\mathcal{T}\partial_{\tilde{z}} \right) \xi(\tilde{r},\tilde{\tau}),
\end{align}
where 
\begin{align}
a_{ij}=\mathcal{C}\hat{n}_{i}\hat{n}_{j}+\hat{n}_{(i}\mathcal{P}_{j)k}\mathcal{B}^{k}+\mathcal{P}_{ik}\mathcal{P}_{jl}\mathcal{A}^{kl}.
\end{align}
Since $\mathcal{C}, \mathcal{B}^{k},  \mathcal{A}^{kl}$ and $\mathcal{T}$ do not change by the pure gradients, the observed large scale power spectrum is not modified. 

\section{discussion and conclusion}
Our aim in this paper is two fold. First, we emphasize  that the effect of a gradient mode on  observables should vanish, providing that the derivation is consistent. In this manner, we extend the investigation of the long mode modulations to other large scale observables taking into account projection effects that are inherent in a consistent relativistic formulation of the large scale observables. In the relativistic formulation,  defining observables is crucial. These Observables are gauge invariant quantities made of information on the past light cone. The measured value of these observables is different because of the inhomogeneity effects. As an example, the cosmic rulers are one observable with a known physical scale like the BAO. Their observed scale can be compared to their physical scale.
The observables in the cosmic rulers can be presented in terms of  a scalar, a vector and a tensor quantity. 
The relativistic formulation of observables yields that the effects of gradient modes vanishes on all cosmic ruler observables.
The other important quantity in the relativistic formulation is the galaxy number counts.
Additionally, we show that  different projection effects cancel each other as well, using the relativistic galaxy number counts.\\

Second, we emphasize that the effects of a gradient mode on the bias vanishes in a consistent bias formulation. It is important to note this is possible even though there is a coupling of gradient modes to short modes. That is non-gaussian corrections induced by gravitational coupling vanish. Since the halo number density depends on the matter variance, we first derived a consistency condition for how the matter power spectrum is modulated in the presence of a long gradient mode. As halo formation is a local phenomena, we used local coordinates to define the matter  variance. This yields that the matter variance does not change in local coordinates. This confirms that the definition (\ref{halo}) is consistent with equivalence principle.\\

The vanishing effect of gradient modes is favored by the equivalence principle.
It is shown that the local physical effects of  long mode perturbations start at $(k_{L}x)^2$ order if one uses conformal Fermi coordinates as implied by the equivalence principle \cite{Dai:2015rda}. As a result,  the pure gradient mode does not contribute to the tidal term at linear order. However, to calculate the observables at late times, we have to transform from the conformal Fermi coordinates to the observed coordinates. The gradient mode should vanish in a consistent  general relativistic formulation.\\


{\bf Acknowledgments:}\\

We would like to thank Hassan Firouzjahi, Ali Akbar Abolhasani and Reza Mansouri for useful discussions and comments.\\

\end{document}